\documentclass{aa}
\usepackage{epsfig}
\def\ltsima{$\; \buildrel < \over \sim \;$}
\def\simlt{\lower.5ex\hbox{\ltsima}}
\def\gtsima{$\; \buildrel > \over \sim \;$}
\def\simgt{\lower.5ex\hbox{\gtsima}}
\begin{document}
\thesaurus{}
   \thesaurus{3(11.09.1: NGC4593;  
		11.14.1; 
		11.19.1; 
		13.25.2)} 

\title{The complex 0.1-200~keV spectrum of the Seyfert 1 Galaxy NGC~4593}
\author{M. Guainazzi\inst{1}, G.C. Perola\inst{2}, G.Matt\inst{2}, F.Nicastro\inst{3,2}, L.Bassani\inst{4}, F.Fiore\inst{3,5}, D.Dal Fiume\inst{4}, L.Piro\inst{6}}

\institute{
{Astrophysics Division, Space Science Department of ESA, ESTEC, Postbus 299,
NL-2200 AG Noordwijk, The Netherlands}
\and
{Dipartimento di Fisica ``E.Amaldi'', Universit\`a di Roma 3, Via della Vasca 
Navale 84, I-00146 Roma, Italy}
\and
{Osservatorio Astronomico di Roma, Via dell'Osservatorio 5, I-00040 Monteporzio Catone, Italy}
\and
{Istituto Tecnologie e Studio Radiazioni Extraterrestri, CNR, Via Gobetti 101, I-40129 Bologna, Italy}
\and
{BeppoSAX Science Data Center, Via Corcolle 19, I-00131 Roma, Italy}
\and
{Istituto di Astrofisica Spaziale, CNR, Via Fosso del Cavaliere, I-00133 Roma, Italy}
}
   
\offprints{M.Guainazzi (mguainaz@astro.estec.esa.nl)}

\date{Received 28 September 1998; accepted 17 March 1999}

\maketitle

\markboth{M.Guainazzi et al.}{The complex 0.1-200~keV spectrum of the 
Seyfert 1 Galaxy NGC~4593}

\begin{abstract}

We report on the first observation of the Seyfert~1 galaxy NGC~4593 in the
0.1--200~keV band, performed with the BeppoSAX observatory.
Its spectral components are for the first time
{\it simultaneously} measured: a power-law with photon spectral index
$\Gamma \simeq$~1.9; the Compton-reflection of the primary power-law;
a moderately broad
(${\rm \sigma > 60}$~eV) K$_{\alpha}$ fluorescent line from neutral iron;
and an absorption edge, whose threshold energy is consistent with
K-shell photoionization from O{\sc vii}. The amount of reflection
and the iron line properties are consistent with both being produced
in a plane-parallel, X-ray illuminated relativistic accretion disc surrounding
the nuclear black hole, seen at an inclination of ${\rm \simeq 30^{\circ}}$.
Any cutoff of the intrinsic continuum is constrained to lay above
150~keV.
The claim for a strongly variable soft excess is dismissed
by our data and by a reanalysis of archival ASCA and ROSAT data.

\end{abstract}

  \keywords   {Galaxies: individual: NGC4593 --
		Galaxies: nuclei --
		Galaxies: Seyfert --
		X-rays: general
		}

\section{Introduction}

It is widely accepted that the huge energy output of
Active Galactic Nuclei (AGN) is due to the release of
gravitational energy by matter falling onto a supermassive
($\sim 10^6$--${\rm 10^8 M_{\odot}}$)
black hole.
It has been known since the beginning of the X-ray astronomy
that the spectrum
of Seyfert~1 galaxies above a few keV and up to a few tens of keV
can be described at the 0-th order
by a simple power-law, with typical photon index $\Gamma \simeq
1.5-2$ (Mushotzky 1984). Later on,
an emission line from neutral or mildly ionized iron (Holt et al. 1980;
Perola et al. 1986; Pounds et al. 1990) and
a flattening of the spectrum above $\sim 10$~keV
(Pounds et al. 1990; Piro et al. 1990)
were discovered to be
common spectral features in the class as well. Both these features
have been interpreted as the effect of reprocessing of the primary
radiation by optically thick matter surrounding the nucleus
(George \& Fabian 1991; Matt et al. 1991).
The detection of the double-horned and redshifted profile of the iron
line in a long-look ASCA observation of the archetypical Seyfert~1
MCG-6-30-15 (Tanaka et al. 1995; Iwasawa et al. 1996), together with
the general evidence that the iron lines are on the average broad in this
class (Nandra et al. 1997), gave the first direct evidence that the
reprocessing matter is located close to
the black hole, probably in a Keplerian accretion disk.

In soft X-rays, an excess above the extrapolation of the
high-energy power-law was
measured in 50\% of the Seyferts observed by EXOSAT
(Turner \& Pounds 1989) and about 90\% of those observed
by ROSAT (Turner et al. 1993; Walter \& Fink 1993).
The good correlation with Optical/UV
(Walter \& Fink 1993; Puchnarewicz et al. 1996; Laor et al. 1997)
has traditionally supported the idea that this excess represents
the hard tail of thermal emission from the accretion
disk (Czerny \& Elvis 1987), although in at least a few
cases reprocessing by ionized matter seems a
more viable explanation (Piro et al. 1997; Guainazzi et al. 1998b).
However, caution must be employed when evaluating the soft excess
in band-limited detectors and/or from multi-instrumental fits of not
simultaneous observations.
The extrapolation of 1--10~keV power-law fits can yield ``faked''
soft excesses, because the so determined spectral indices can be
harder than the true one,
due to the flattening contribution of the
reflection component. Eventually,
the discovery of absorption edges from
highly ionized species of oxygen in almost 50\% of the Seyfert~1s
observed by ASCA so far (Reynolds 1997; George et al. 1998) has
suggested that in some cases the apparent soft excess
could be simply due to a mis-fit of a complex and ionized absorber.

Given the spectral complexity outlined above
a detailed and self-consistent
description of the X-ray spectra of Seyfert~1s requires broadband
spectral observations.
Moreover, the various spectral components are expected to be
produced in different physical regions
(see Mushotzky et al. 1993 for a review of the pre-ASCA
interpretative scenario, which is still largely valid)
and are therefore expected to exhibit different variability
patterns and/or a delayed response to the changes of the
nuclear radiation flux.
The simultaneous measure of all the spectral components is
a crucial requirement for any X-ray observation of
Seyfert galaxies (see the discussion in Cappi et al. 1996
about the dependence of the iron line measurements on
the underlying continuum determination).
The Italian-Dutch satellite BeppoSAX (Boella et al. 1997a)
carries a scientific payload which covers
the unprecedented wide energy band between 0.1 and 200~keV, with
imaging capabilities and good energy resolution in the 0.1--10~keV
band. A program of spectral survey of a sizeable sample of
Seyfert~1 galaxies is ongoing, and in this {\it paper} we report the
results of the observation of NGC~4593. 

NGC~4593 (${\rm \alpha_{2000}=12^h 39^m 39^s.4}$,
${\rm \delta_{2000}=-5^{\circ} 20' 39''}$) is a nearby (${\rm z = 0.009}$) barred spiral galaxy
of Hubble type SBb, which hosts a Seyfert 1 nucleus. The
Spectral Energy Distribution (SED) is characterized by a very weak or missing
``blue bump'' (Santos--Lle\'o et al. 1994). In X--rays,
the source is variable, on timescales of weeks-months, by
a factor of $\sim$~3 in the intermediate ({\it i.e.}:
2--10~keV) and $\sim$~4-5 in the soft X--rays
(Ghosh \& Soundararajaperumal 1993). In NGC~4593 a soft
excess above the extrapolation of the intermediate X--ray
power-law was observed both by EXOSAT (Ghosh \& Soundararajaperumal 1993;
Santos--Lle\'o at al. 1995) and ROSAT (Walter \& Fink 1993).
It was claimed to be variable, ranging from 0 to $\simeq 270\%$
of the intermediate X-ray extrapolated flux.
However, this picture is further complicated by the ASCA discovery of a warm
absorber (Reynolds 1997),
with optical depths of the O{\sc vii} and O{\sc viii} photoionization
edges equal to $\simeq 0.3$ and $0.1$, respectively. There was no
significant evidence of broadening of a weak (Equivalent Width
${\rm EW \simeq 90}$~eV)
fluorescent line from neutral iron
(centroid energy ${\rm E_c \simeq 6.35}$~keV) in the ASCA data
(Nandra et al. 1997). In the hard X-ray regime, BATSE measured
a 20--100~keV flux of $(9.8 \pm 1.6) \times 10^{-11}$~erg~cm$^{-2}$~s$^{-1}$
(Malizia et al. 1997),
whereas the 50--150~keV OSSE 2-$\sigma$ upper limit was $4.4 \times
10^{-11}$~erg~cm$^{-2}$~s$^{-1}$ (Johnson et al. 1993).

This {\it paper} is organized as follows. In Sect.~2 we describe
the data reduction and preparation. Sect.~3 and 4 deal with
the timing and spectral analysis of BeppoSAX data, respectively.
In Sect.~5, we compare our findings with a reanalysis of
archival ASCA data of
the same object. The results
are discussed in Sect.~6.

\section{Data Reduction}

The Italian-Dutch satellite BeppoSAX carries
four co-aligned Narrow Field Instruments. Two of them are gas scintillation
proportional counters with imaging capabilities: the Low Energy
Concentrator Spectrometer (LECS, 0.1--10~keV, Parmar et al. 1997)
and the Medium Energy Concentrator Spectrometer (MECS, 1.8--10.5~keV,
Boella et al. 1997b). The other two instruments are direct-view
detectors, seen through a rocking collimator to achieve
a continuous monitoring of the background: the High Pressure Gas
Scintillator Proportional Counter (HPGSPC, 4-120~keV, Manzo et al. 1997)
and the Phoswitch Detector System (PDS, 13-200~keV, Frontera et al. 1997).
The HPGSPC is
tuned for spectroscopy of very bright sources with good energy resolution,
while the PDS possesses an unprecedented sensitivity in its energy
bandpass. Only LECS, MECS and PDS data will be considered in this {\it paper}.
The HPGSPC data points could not in fact provide significant constraints on
the spectral shape of NGC~4593.

NGC4593 was observed by BeppoSAX from 1997 December 31 05:30:18 UT to 1998
January 2 07:10:17 UT. Data were telemetred in direct modes for all
instruments, which provide full information about the arrival time,
energy, burst length/rise time (RT) and, when available, position for each
photon. 
Standard reduction procedures and screening criteria have been
adopted to produce linearized and equalized event files. In particular,
time intervals have been excluded from the scientific product accumulation
when: the angle between the pointing direction and the Earth's
limb was $< 5^{\circ}$; the momentum associated to Geomagnetic Cutoff
Rigidity was $> 6$~GeV/c; the satellite passed through the South Atlantic
Geomagnetic Anomaly (SAGA). The PDS data have been further screened by
eliminating 5 minutes after any SAGA passage to avoid gain instabilities
due to the recovery to the nominal voltage value after instrumental
switch-on. The RT selection has been performed using crystal
temperature dependent thresholds (instead of the fixed thresholds in the
standard processing). Total exposure times were
31426~s, 84195~s,
and 38467~s for the LECS, MECS and PDS,
respectively.

Spectra of the imaging instruments
have been extracted from circular regions
of radius 8' and 6' around the apparent centroid of the source for the
LECS and MECS, respectively. The
background subtraction has been performed
using spectra extracted from blank sky event files in the same region
as the source. The
background contributes 5\% and 3\% to the LECS and MECS count rates
in the relevant energy band passes. PDS background-subtracted
products (spectra and light curves) have been produced by plain subtraction
of the ``off-'' from the ``on-source'' products. Total net count rates
were $0.310 \pm 0.003$~s$^{-1}$, $0.350 \pm 0.002$~s$^{-1}$ and
$0.80 \pm 0.04$~s$^{-1}$ in the LECS, MECS and PDS, respectively.

In this {\it paper}: errors are quoted at 90\% level of confidence for two
interesting parameters (${\rm \Delta \chi^2 = 4.61}$,
Lampton et al. 1976); energies are in the source rest frame;
the cosmological parameters $q_0 = 0.5$ and $H_0 = 50$~km~s$^{-1}$~Mpc$^{-1}$
are assumed, unless otherwise specified.

\section{Timing analysis}

\begin{figure}
\begin{center}
\epsfig{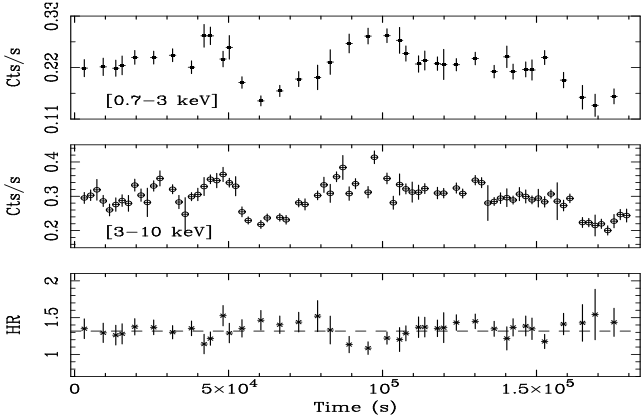}
\end{center}
\caption{Light curves in the 0.7--3~keV (LECS, upper panel)
and 3--10~keV (MECS, central panel) energy bands
and the corresponding HR (lower panel).
The binning time is 2048~s. The light curves are not background-subtracted}
\label{fig1}
\end{figure}

\begin{figure}
\begin{center}
\epsfig{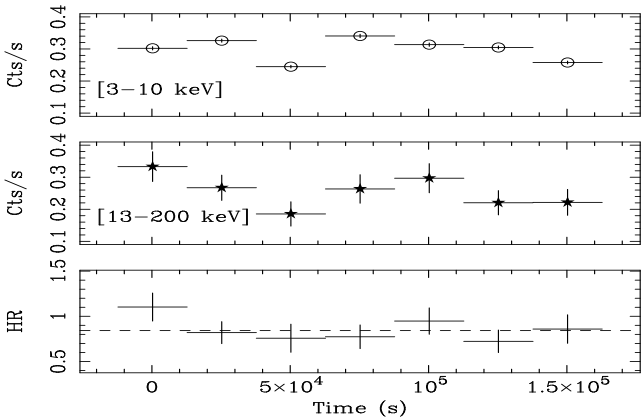}
\end{center}
\caption{Light curves in the 3--10~keV (MECS. upper panel),
13-200~keV (PDS, central panel) energy bands
and the corresponding HR (lower panel).
The binning time is 25000~s. The MECS light curve is not background-subtracted}
\label{fig2}
\end{figure}

\begin{figure}
\begin{center}
\epsfig{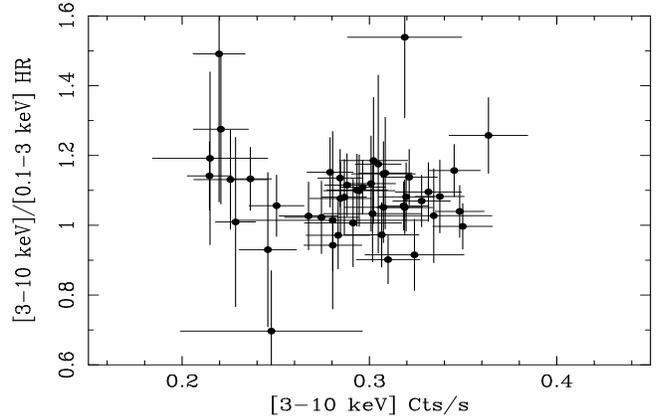}
\end{center}
\caption{0.1--3 vs. 3--10~keV HR versus the 3--10~keV count rate.
Each data point corresponds to a 2048~s light curve bin}
\label{fig10}
\end{figure}

NGC~4593 is well known to display rapid fluctuations in X-rays, with
two-folding time $\simeq 1.1$~h (Barr et al. 1987).
The count rates in the BeppoSAX instruments span indeed a dynamical
range of about a factor of 2 on timescales as low as
$3 \times 10^4$~s. We have therefore extensively searched
for spectral variability associated with the flux changes, extracting
light curves and Hardness Ratios (HR) in different energy bands and
with various binning times. In Fig.~\ref{fig1}
and \ref{fig2} we show the light curves in the 0.7--3~keV, 3--10~keV
and 13--200~keV with the corresponding contiguous HR.
The bands have been chosen in order to
sample energy intervals where different spectral components dominate,
see Sect.~4. In particular, the softer band is affected
by the warm absorber, which is basically transparent for photons
of energy higher than 3~keV.
No significant variability of the HR is detected.
If a fit with a constant is performed on
the 3--10~keV vs. 0.7--3~keV HR,
the $\chi^2$ is 29.3 for 38 degrees of freedom (dof).
The same quantity for the 13--200~keV vs. 3--10~keV HR
is 4.8/6~dof.
A tiny perturbation of the intermediate X-ray flux
in a $2 \times 10^4$~s interval around $9 \times 10^4$~s
after the beginning of the observation
is not accompanied by a similar phenomenology
in the soft X--rays and yields therefore a decrease of the HR by
a factor $\simeq 15\%$. However, no
variation in the spectral parameters above the statistical uncertainties
has been detected
in time-resolved spectroscopy (assuming the baseline model of Sect.~4.1).
Similar results are obtained if the bands 0.1--3 and 3--10~keV
are considered (see Fig.~\ref{fig10}).
We will therefore focus in the following on the
time-averaged spectral behavior.

\section{Spectral analysis}

The spectra of the imaging instruments
have been rebinned in order to sample the intrinsic energy
resolution of the detectors with 3 (LECS) or 4 (MECS) energy channels.
Each channel has at least 30 counts, which ensures the
applicability of $\chi^2$ test. The PDS spectrum has been
quasi-logarithmically rebinned, in order to have 16 energy
channels in the 14--200~keV band. The spectra of the
three detectors have been fitted simultaneously. 
Numerical relative normalization factors among the BeppoSAX
instruments have been added to all the following spectral fits.
The reasons is two-fold: a) the BeppoSAX instrument response matrices
employed in this {\it paper} (September 1997 release) exhibit slight
mismatches in the absolute flux calibration; b) the sampling
of the instruments is not strictly simultaneous, due to
the need for operating the LECS only during satellite nights or
the different data selection criteria between imaging instruments
and the PDS. This can affect the flux measured in variable
sources as NGC~4593. 
The LECS to MECS factor has been left free in the fitting
procedure, and turns out to be comprised in the
range 0.74--0.76, which is consistent with typical values observed
so far (0.7--1.1: Grandi et al. 1997; Haardt et al. 1998; Cusumano et
al. 1998).
The PDS to MECS factor has been instead held fixed to 0.8,
as the available statistics was not good enough to provide
independent constraints on it.
This value corresponds to
the multiplication of the best fiducial value
estimated by Cusumano et al. (1998)
by 0.82, to account for the effect of the PDS RT selection algorithm employed
(see Sect.~1). The systematic uncertainty on this parameter can
be estimated $\pm 10\%$. Spectral fits have been performed in the
0.1--4~keV (LECS), 1.8--10.5 (MECS), 14--200~keV (PDS) energy
bands.

\subsection{Continuum shape}

\begin{figure}
\begin{center}
\epsfig{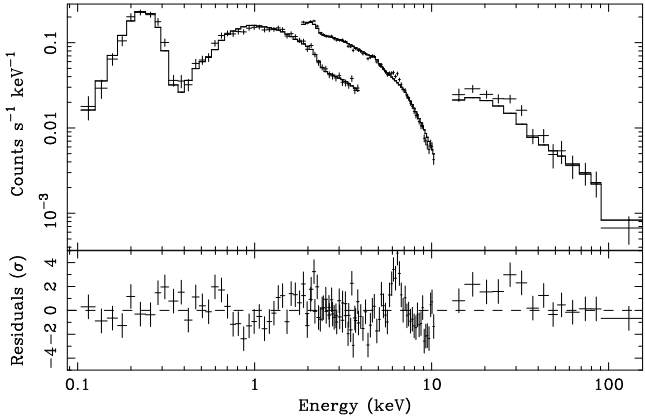}
\end{center}
\caption{Spectra and best-fit model (upper panel) and residuals
in units of standard deviations (lower panel) when a simple
power-law model with photoelectric absorption is applied}
\label{fig3}
\end{figure}

In Fig.~\ref{fig3} the result is shown, when a simple power-law model
with photoelectric absorption is applied. The quality of the fit is
rather poor ($\chi^2 = 262.6/122$~dof). The main deviations are due
to: (a) an absorption feature starting at $E \simeq 0.7$~keV; (b)
a prominent emission line with centroid energy $E \simeq 6.5$~keV;
(c) a ``bump'' in the PDS band, peaking at energy $E \simeq 25$~keV.
The residuals between 0.3 and 0.6~keV
(immediately red wards the absorption feature) are systematically
positive. In principle, this may be due to the emergence of a soft excess
in this energy range. However, the recovery of the residuals to values
consistent with zero below 0.3~keV leads to the suggestion the the
above feature is simply the typical wavy residual produced by a
mis-fit absorption edge (see {\it e.g.} Nandra \& Pounds 1992).
The BeppoSAX broadband allows to study simultaneously 
the whole spectral complexities expected on the basis of previous
band-limited
measures of this object and of the Seyfert~1s as a class (cf.
Sect.~1).
We have then defined a ``baseline'' model, where a Compton
reflection component from a neutral slab
(model {\verb!pexrav!} in {\sc Xspec}, Magdziarz \& Zdziarski 1995), an
emission line and a photoionization absorption edge are superposed
to the photoelectric absorbed power--law.
The possibility that the reflecting matter is substantially
ionized is not required by the data. The model depends on the
heavy element abundance (which has been held fixed to the solar
one), and on the angle between the
normal to the slab and the line of sight (``inclination angle''
$\theta$).
$\theta$ has been
held fixed to $30^{\circ}$ hereinafter (the best-fit
value arising from a fit of the iron line profile with a relativistic
model, see Sect.~4.2).
The only free parameter in addition to the power-law model
of the continuum is the relative normalization ${\rm R}$ between the
reflected and the primary components (equal to 1 for an isotropic
source, illuminating a plane-parallel infinite slab).
The best-fit
${\rm R}$ and $\Gamma$ are basically unaffected, if
one assumes that the reflector
is seen face-on ({\it i.e.}: ${\rm \theta = 0^{\circ}}$).
The ``baseline'' model yields a
very good $\chi^2 = 99.0/116$~dof.
The Table~\ref{tab1} reports the best-fit parameters and
results.
\begin{table*}
\caption{BeppoSAX best-fit results. {\it wa}~=~photoelectric absorption from
neutral matter; {\it po}~=~power-law; {\it px}~=~Compton reflection;
{\it ga}~=~Gaussian line; {\it ed}~=~photoionization absorption edge;
{\it bk}~=~broken power-law}
\begin{tabular}{lccccccccc} \hline
Model & ${\rm N_H}$ & $\Gamma$ & ${\rm R}$ & ${\rm E_c}$ & ${\rm \sigma}$ & 
${\rm EW}$ & ${\rm E_{th}}$ or ${\rm E_{break}}$ & ${\rm \tau}$ or ${\rm \Gamma_{soft}}$ & ${\rm \chi^2/}$~dof \\ 
& (${\rm 10^{20}}$~cm$^{-2}$) & & & (keV) & (keV) & (eV) & (keV) & & \\ \hline
{\verb!wa*po!} & $1.55 \pm^{0.14}_{0.13}$ & $1.70 \pm 0.04$ & & & & & & & 
262.6/122 \\
{\verb!wa*px!} & $1.9 \pm 0.2$ & $1.79 \pm 0.04$ & $0.9 \pm^{0.4}_{0.3}$ & & & 
& & & 183.9/121 \\
{\verb!wa*(px+ga)!} & $1.9 \pm 0.2$ & $1.80 \pm 0.04$ & $0.8 \pm^{0.4}_{0.3}$ 
& $6.41 \pm^{0.14}_{0.13}$ & $0.2 \pm^{0.3}_{0.2}$ & $170 \pm^{70}_{60}$ & & & 
134.8/118 \\
{\verb!wa*ed*(px+ga)!}$^a$ & $2.3 \pm 0.3$ & $1.87 \pm 0.05$ & $1.1 \pm 0.4$ & 
$6.42 \pm^{0.17}_{0.14}$ & $0.3 \pm^{0.3}_{0.2}$ & $190 \pm^{90}_{60}$ & $0.77 
\pm^{0.06}_{0.05}$ & $0.43 \pm^{0.17}_{0.16}$ & 98.5/116 \\
{\verb!wa*ed*(px+ga)!}$^b$ & $2.3 \pm 0.3$ & $1.86 \pm 0.05$ & $1.0 
\pm^{0.6}_{0.4}$ & $6.42 \pm^{0.15}_{0.14}$ & $0.3 \pm^{0.3}_{0.2}$ & $200
\pm^{100}_{70}$ & $0.78 \pm 0.06$ & $0.42 \pm^{0.17}_{0.16}$ & 99.0/116 \\ 
{\verb!wa*ed*(px+ga)!}$^c$ & $1.5 \pm^{1.7}_{1.4}$ & $1.96 \pm^{0.05}_{0.03}$ 
& $0.6 \pm^{1.0}_{0.0}$ & $6.5 \pm^{0.3}_{0.2}$ & $0.5\pm^{1.0}_{0.4}$ & 
$280\pm^{410}_{140}$ & $0.68 \pm 0.03$ & $0.30\pm^{0.11}_{0.08}$ & 1396.4/1357 
\\  
&  &  &  &  &  &  & $0.83\pm^{0.05}_{0.03}$  & $0.13 \pm^{0.05}_{0.07}$ & \\
{\verb!wa*(bk+px+ga)!} & $1.4 \pm 0.3$ & $1.87 \pm^{0.13}_{0.06}$ & $1.2 \pm^{1.5}_{0.5}$ & $6.42 \pm^{0.18}_{0.17}$ & $0.4 \pm^{0.6}_{0.3}$ & $240 \pm^{170}_{90}$ & $2.2 \pm 0.5$ & $1.67 \pm 0.07$ & 133.6/116 \\ \hline
\end{tabular}
\noindent
$^a$${\rm \theta=0}$; \\
$^b$${\rm \theta=30^{\circ}}$, ``baseline model'' in text. \\
$^c$fit on the ASCA data (see Sect.5)
\label{tab1}
\end{table*}

\begin{figure*}
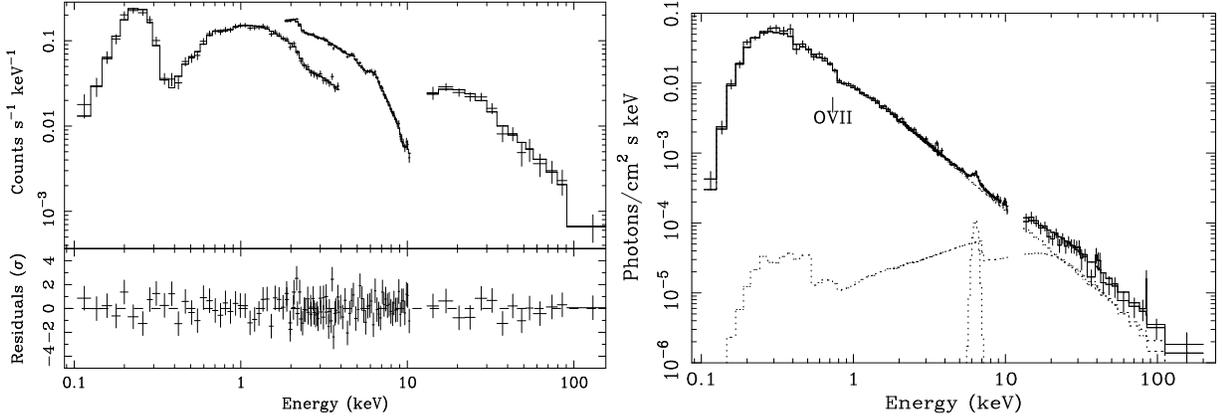

\begin{center}
\epsfig{figure=H1211.f5,height=8.0cm,width=5.5cm,angle=-90}
\epsfig{figure=H1211.f6,height=8.0cm,width=5.5cm,angle=-90}
\end{center}
\caption{Left: spectra and best-fit model (upper panel) and residuals
in units of standard deviations (lower panel) when the ``baseline'' model
is applied. Right: unfolded energy spectrum and best-fit model (solid
lines). The direct and reflected continuum and the emission line
are
separately indicated with dotted lines. The location of the {\sc O vii}
photoionization absorption edge in the observers frame is labeled}
\label{fig5}
\end{figure*}

It is worth noticing that no soft excess above the extrapolation
of the high-energy power--law is required.
A model, where one replaces the absorption edge with a
continuous break of the primary power-law provides a much worse fit
($\chi^2 = 133.6/116$~dof) and leaves significant residuals in the
0.2-1.3~keV energy range. Moreover, the best-fit spectrum is
convex ({\it i.e.}: ${\rm \Gamma_{soft} < \Gamma}$), further
demonstrating that no soft excess is present in the data.
The inferred absorbing column
density is consistent with the Galactic contribution
along the line of sight as measured by Elvis et al. (1989,
${\rm N_{H_{Gal}} = 1.97 \times 10^{20}}$~cm$^{-2}$).
The best--fit model and
deconvolved spectra are shown in
Fig.~\ref{fig5}.
The 0.1--2~keV, 2--10~keV and 20--100~keV fluxes are $\simeq 2.11 \times
10^{-11}$, $3.74 \times 10^{-11}$ and
$7.22 \times 10^{-11}$~erg~cm$^{-2}$~s$^{-1}$, respectively. They correspond
to rest frame luminosities of $5.80 \times 10^{42}$, $1.03 \times 10^{43}$
and $1.99 \times 10^{43}$~erg~s$^{-1}$, respectively.

\begin{figure}
\begin{center}
\epsfig{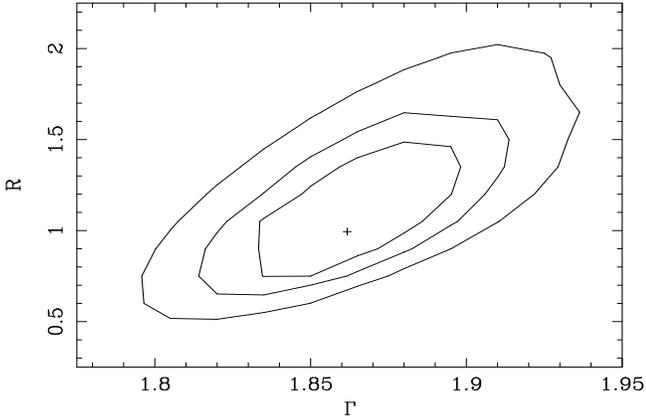}
\end{center}
\caption{Contour plot of the relative normalization between the
primary and Compton reflected spectral components versus the
intrinsic photon index, when the
``baseline'' model is employed. Iso-$\chi^2$ curves are at
68\%, 90\% and 99\% confidence
level for two interesting parameters ($\Delta \chi^2 = 2.3$, 4.6 and
9.2)}
\label{fig6}
\end{figure}

\begin{figure}
\begin{center}
\epsfig{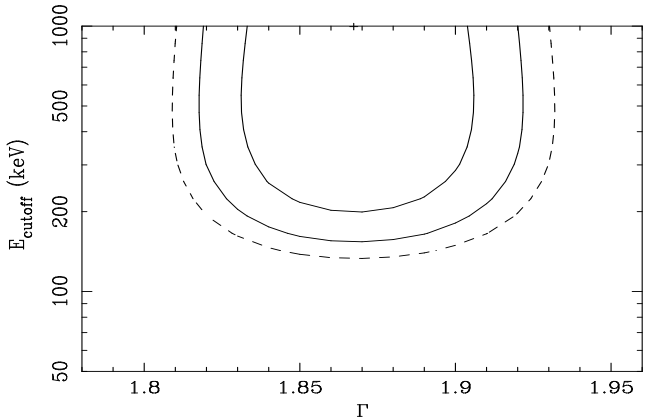}
\end{center}
\caption{Contour plot of the cutoff energy of the primary
power-law component versus the photon index when the
``baseline'' model is employed}
\label{fig7}
\end{figure}

The unprecedented BeppoSAX energy coverage allows the
simultaneous determination of the primary radiation steepness and of the
Compton reflection intensity with the best accuracy ever achieved.
In Fig.~\ref{fig6} the contour plot for the photon index versus
the relative normalization between the reprocessed and primary components
${\rm R}$ is shown.
The latter parameter is 1 if the reflection occurs in an
infinite, plane-parallel slab and the primary source emits isotropically.
Higher values are formally possible, and may be due to the geometry
of the reflecting matter, covering more than $2 \pi$ ({\it e.g.} in
a concave or warped accretion disk), to an intrinsic anisotropy
of the primary source, or to a delayed response of the reflecting
matter to changes of the primary flux.
At 90\% level of confidence for two interesting parameters,
the photon index is constrained between 1.81 and
1.91. R is consistent with a plane-parallel geometry of the reflecting
matter (90\% confidence interval between 0.6 and 1.6).
Any cut-off of the intrinsic power-law is constrained to lay at energies
higher then 150~keV (see Fig.~\ref{fig7}).
It is worth noticing that the statistical relative
uncertainties on $\Gamma$ and ${\rm R}$ are rather small, despite the
strong correlation between the two parameters
(note the strongly inclined contour plot in
Fig.~\ref{fig7}). For comparison,
the single parameter 68\% statistical errors
(${\rm \Delta \Gamma = 0.02}$ and ${\rm \Delta R/R = 0.17}$)
are 4.5 and 3.5 times smaller than the ones
obtained from the Ginga observations of the same Seyfert~1
(Nandra \& Pounds 1994). It should however be noticed
that the residual systematic uncertainties on the relative PDS to MECS
normalization factor affect the accuracy of both these parameters,
with additional uncertainties of about  1\% and 30\%, respectively.

\subsection{On the K$_{\alpha}$ iron line}

\begin{figure}
\begin{center}
\epsfig{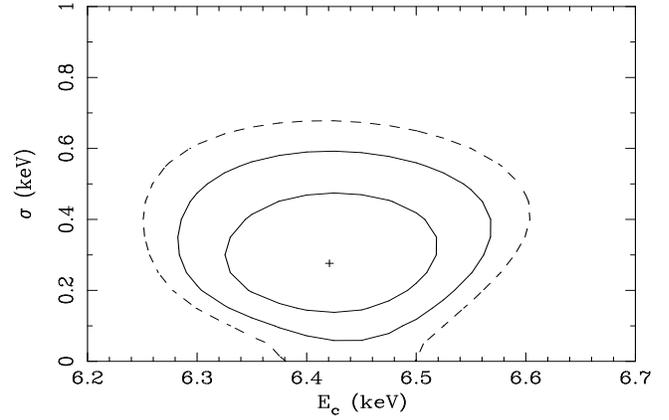}
\end{center}
\caption{Contour plot of the intrinsic width versus
the centroid energy of the iron line, when the ``baseline''
model is employed}
\label{fig8}
\end{figure}

The centroid energy of the iron line is well consistent with
K${\alpha}$ fluorescence from neutral iron.
The line is broad if a simple
broad Gaussian profile is used to describe it (see Fig.~\ref{fig8});
the intrinsic width is
comprised in the range 60--600~eV at 90\%
level of confidence for two interesting parameters
(best-fit value $\simeq 300$~eV). 
We have tried also
alternative parameterization of the line. Adding a further
``narrow'' line ({\it i.e.}: intrinsic width $\sigma$ held fixed at zero)
to the ``baseline'' model
results in no improvement of the quality of the fit
($\Delta \chi^2 = 0$).
Given the intrinsic width of the line and the agreement
between the derived best--fit line EW
and the theoretical expectations if the line is
produced in an X--ray illuminated relativistic accretion disk
(Matt et al. 1992), we have tried to use a self-consistent
model of line emission from a relativistic accretion disk (model
{\verb!diskline!} in {\sc Xspec}, Fabian et al. 1989).
If all its parameters are allowed to be free, most of them
are totally unconstrained.
No further constraint comes from the ASCA
data as well (Nandra et al. 1997). If
it is assumed that
the inner radius ${\rm R_i}$ of the emitting region
is 6 gravitational radii
(${\rm R_G \equiv GM/c^2}$);
the emissivity law index is equal to -2.5
(Nandra et al. 1997);
the inclination of the line and Compton reflection continuum emitting region is
the same; and the line is neutral
({\it i.e.}: ${\rm E_c = 6.4}$~keV); then
${\rm \theta = 32 ^{\circ} \pm^{23}_{12}}$,
${\rm \log(R_o) > 2.1}$ and
${\rm EW = 240 \pm^{70}_{60}}$~eV.
The quality of the iron line modeling (${\rm \chi^2 =98.9/114}$~dof)
is comparably good as for the broad Gaussian.

\subsection{The warm absorber}

In 12 out of 24 Seyfert galaxies observed by ASCA, absorption
edges from ionized species of oxygen have been detected (Reynolds
1997), which have been interpreted as the imprinting of warm gas along
the path from the nucleus and us. NGC~4593 is one of these
objects, and the BeppoSAX observation confirms this outcome,
thanks to the detection of an absorption edge with threshold energy
${\rm E_{th} = 0.78 \pm 0.06}$~keV.
The edge energy is consistent with the K-photoionization
threshold energy of O{\sc vii}. 
We have tried to give a
qualitative characterization of the ionization and chemical structure
of the absorbing matter, by tentatively including in the fit four absorption
edges, with threshold energies held fixed to the values expected from
O{\sc vii} (0.739~keV), O{\sc viii} (0.871~keV), Ne{\sc ix}
(1.196~keV) and Ne{\sc x} (1.362~keV). However, the available statistics
is not good enough to give us significant constraints. Only the
O{\sc vii} edge yields a significant detection
(${\rm \tau_{OVII} = 0.32^{+0.17}_{-0.16}}$), while only upper limits can be
obtained for the other three edges (${\rm \tau_{OVIII} < 0.22}$,
${\rm \tau_{NeIX} < 0.05}$, ${\rm \tau_{NeX} < 0.03}$). Using
{\sc Cloudy} (Ferland 1996), we have constructed  a grid of self-consistent
models of the spectra transmitted through a ionized gas in thermal
and ionization equilibrium, when the SED is the one
observed in NGC~4593 (Santos-Lle\'o et al. 1995). They depend on the
ionization parameter ${\rm U}$ (defined
as the dimensionless ratio between the number of Hydrogen
ionizing photons and the electron density of the gas),
and the warm column density ${\rm N_W}$. The
fit is comparably good as the phenomenological description of the
``baseline'' model ($\chi^2 = 104.6/116$). The best-fit parameters of
the warm absorbing matter are
${\rm N_W = (2.5\pm^{1.5}_{1.2}) \times 10^{21}}$~cm$^{-2}$ and ${\rm \log(U) =
-0.7 \pm 0.3}$.
For a source with ${\rm \alpha_{ox} \simeq 1.20}$, the best-fit ${\rm U}$
corresponds to a value of the more common
${\rm \xi \equiv L/n^2 R \simeq}$ a few. 
The continuum parameters are slightly affected
(${\rm \Gamma = 1.92 \pm 0.10}$; ${\rm R = 1.3 \pm
0.5}$), but remain consistent with the one of the ``baseline'' model
within the statistical uncertainties.

\section{Comparison with ASCA and ROSAT results}

\begin{figure}
\begin{center}
\epsfig{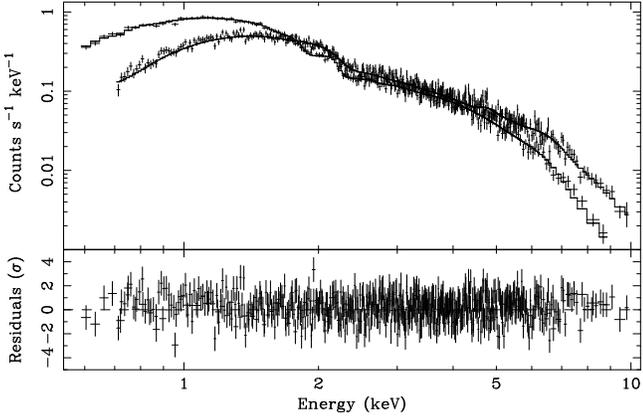}
\end{center}
\caption{Spectra and best-fit model (upper panel) and residuals in units
of standard deviations (lower panel) when the modified BeppoSAX ``baseline''
model is applied to the January 1994 NGC~4593 ASCA observation. Only
SIS1 and GIS3 data are shown for clarity}
\label{fig9}
\end{figure}

Nandra et al. (1997) detected with ASCA
a weak (${\rm EW = 90\pm^{40}_{30}}$~eV)
and narrow iron line, whose best-fit parameters are inconsistent
with the ones corresponding to the ``baseline'' BeppoSAX best-fit
model. We have therefore re-analyzed the ASCA observation to
check the robustness of the iron line variability.
ASCA (Tanaka et al. 1994)
payload include a pair of Charged-coupled devices (SIS0 and SIS1, 0.57--9~keV)
and a pair of gas scintillation proportional counters
(GIS2 and GIS3, 0.7--10~keV).
It observed NGC~4593 on January 9, 1994.
Spectra have been extracted from the screened event files, which have
been filtered according to standard criteria, using extraction radii
of 2.30', 3' and 6' for the SIS1, SIS0 and GIS, respectively.
Net exposure times amount to 30 and 33~ks for the SIS and GIS, respectively.
The spectra
have been rebinned in order to have at least 20 counts per energy channel.
Background
spectra have been extracted from blank sky pointing event files, using the
same extraction region in detector coordinates as the source. The same
BeppoSAX ``baseline'' model has been applied to the spectra of
all detectors simultaneously, except for the fact that
ASCA is capable to resolve the O{\sc vii} and
O{\sc viii} absorption edges. The parameter ${\rm R}$, which cannot
be constrained by ASCA given the limited
bandwidth of its instruments, has been forced to be comprised in the 90\%
confidence level range determined by BeppoSAX data, {\it i.e.}: 0.6--1.6.
The time-averaged 2--10~keV flux in the ASCA observation is very close to that
measured by BeppoSAX
(${\rm F \simeq 3.70 \times 10^{-11}}$~erg~s$^{-1}$~cm$^{-2}$).
No significant deviation exists between the ASCA and BeppoSAX
best-fit parameters, with the exception of a slight steepening of the intrinsic
power-law (${\rm \Delta \Gamma = 0.10 \pm^{0.10}_{0.07}}$, cf.
Table~\ref{tab1}). In particular, the iron emission line best-fit
parameters are consistent with the one obtained from the BeppoSAX data
analysis within the statistical uncertainties and there is no evidence
for a much weaker line in the ASCA data.
It is worth noticing that no soft excess is required.
The ASCA spectra and residuals against the best-fit are shown
in Fig.~\ref{fig9}.

Walter \& Fink (1993) report the measure of
a very steep (${\rm \Gamma = 2.5}$)
intrinsic spectral
index with ROSAT. However, their analysis does not take into account the
possible contribution of
a ionized absorber. We therefore reanalyzed the data of a
pointed NGC~4593 ROSAT/PSPC observation,
to compare BeppoSAX and ROSAT findings.
The ROSAT observation was performed July 14 1992. 
The average spectrum was extracted,
using a circular region about 2' radius around the source centroid.
The background spectrum was extracted
from a surrounding annulus of radii 4'.3 and
7'.5, after removing a 2' circular area around a weak
serendipitous source at ${\rm \alpha_{2000} = 12^h 39^m 39^s.2}$,
${\rm \delta_{2000} = -5^{\circ} 26' 13''}$. Total exposure
time is 1261~s. Publicly available
response matrices, appropriate for the epoch of the observation, were
employed and spectral fit were performed in the energy range 0.1--2~keV,
where these matrices are best calibrated.
A simple power-law model with photoelectric absorption is an adequate
description of the spectrum (${\rm \chi^2 = 9/17}$~dof), with best-fit
parameters: ${\rm N_H = (2.1 \pm 0.5) \times 10^{20}}$~cm$^{-2}$,
${\rm \Gamma = 2.13 \pm^{0.19}_{0.17}}$, 0.1--2~keV flux $\simeq 1.26
\times 10^{-11}$~erg~cm$^{-2}$~s$^{-1}$. The intrinsic spectral index
derived by ROSAT is indeed steeper than measured by BeppoSAX or ASCA,
but by an amount much smaller than reported by Walter \& Fink (1993).
The soft excess above a ${\rm \Gamma = 1.86}$ power-law is only
20\% in flux against the 270\% reported by these authors.
The different energy band-passes in which the spectra are taken, or
slight residuals misalignment in the cross calibration between
ROSAT detectors and other missions (cf. Yaqoob et al. 1994;
Iwasawa et al. 1998; Iwasawa et al. 1999) may easily account for this effect.
Interestingly enough, no absorption edge is required by the ROSAT data
despite the better effective area of the PSPC in comparison to the
LECS (the energy resolution is comparable at 0.5~keV).
The upper limit on the optical depth of a 0.77~keV absorption edge is
0.26. This might be suggestive of a long-term {\it positive}
correlation between flux and warm absorber features, since the 0.1--2~keV
flux is 60\% in the ROSAT observation than in the
BeppoSAX one. This aspect may be worth further investigation by future
X-ray monitoring programs.

\section{Discussion}

The broadband (0.1--200~keV) X-ray spectrum of the Seyfert~1 galaxy NGC~4593
shows most of the typical features of its class. The continuum is
well described by the superposition of a power-law primary component
with $\Gamma \simeq 1.9$ and a Compton reflection component.
The spectral index is broadly consistent both with Ginga (Nandra \&
Pounds 1994) and ASCA (Nandra et al. 1997) measurements, and close to the
typical values observed in Seyfert~1s as a class (Nandra \& Pounds 1994,
Nandra et al. 1997). The best-fit nominal values on ${\rm R}$ is consistent
with a slab geometry for the reflecting matter.
The reflection component is
therefore likely to originate in an X-ray
illuminated relativistic
accretion disk (AD), which subtends a $\simeq 2 \pi$ solid angle from the
nuclear source. This picture is consistent
with the properties of the iron emission line.
Its centroid energy is
well consistent with K${\alpha}$ fluorescence from neutral or
mildly ionized iron (${\rm E_c = 6.42\pm^{0.15}_{0.14}}$). The
line is moderately broad (${\rm \sigma \simeq 0.3}$). Fluorescence
iron lines are expected to be produced, along with the Compton reflection
continuum, in AD and broadened by the combination of gravitational
and Doppler effects if the photons undergo the effects
of the gravitational potential of a supermassive black hole. The
$\simeq 200$~eV EW is around the expectation values if the AD
matter has nearly solar abundances, given
the measured value of ${\rm R}$ (Matt et al. 1992).
NGC~4593 does not suffer the
iron overabundance problem that affects the relativistic
iron line measure in some bright objects (among which
the best studied case so far: MCG-6-30-15, Tanaka et al. 1995;
see also Nandra et al. 1997) or to require a contribution from
a narrow line component, which may originate in the molecular torus
surrounding the nuclear environment in the unification
scenario (Ghisellini et al. 1994; Krolik et al. 1994), and 
observed in several cases (Guainazzi et al. 1996;
Weaver et al. 1997;
George et al. 1998; Guainazzi et al. 1998a). The 
line properties can in principle constrain the
location of the emitting region and the geometry of the AD. 
Fitting the iron line profile with a relativistic model
allows us to constrain the inclination of the system
(${\rm \theta = 32 ^{\circ} \pm^{23}_{12}}$), under the assumptions
that the bulk of the iron line emitting region is neutral.

The combination of intermediate X-ray continuum and iron emission
line properties is hence perfectly consistent with the standard
picture for the production of high-energy radiation
in the nuclear region of radio-quiet nearby AGN.
The primary continuum, with
the ``canonical'' ${\Gamma = 1.9}$ power-law index, is isotropically
produced in the neighborhood of the nuclear supermassive
black hole and reprocessed
via Compton down scattering and fluorescence by an optically thick and
geometrically thin Shakura-Sunyaev disk (Shakura \& Sunyaev 1973),
whose extension is much larger than
the typical size of the primary continuum production region. The reprocessing
matter does not exhibit any substantial ionization or deviation
from cosmic abundances.

EXOSAT (Ghosh \& Soundararajaperumal 1993; Santos-Lle\'o 1994)
observed a strong and remarkably
variable soft excess. Its intensity was between 0 and 45\% of the
extrapolated high-energy flux in seven EXOSAT observations,
when soft and intermediate X-rays were measured simultaneously.
In EXOSAT data, it seems to exist no obvious correlation
between the soft excess and the intermediate X-ray flux or
spectral index (Santos-Lle\'o et al. 1994).
Santos-Lle\'o et al. (1994) discuss the soft excess in terms
of thermal emission from an accretion disc and conclude that
the maximum temperature ${\rm T_{max}}$ should be $\sim 210$~eV.
Such a disk blackbody component would be clearly detectable
by the LECS (if not by the ASCA/SIS).
If we add to the BeppoSAX ``baseline'' model the emission from
a disc blackbody with temperature at the innermost disk radius
${\rm T = 100}$~eV~${\rm = 0.488 T_{max}}$ (Pringle 1981),
its luminosity is $< 3 \times 10^{42}$~erg~s$^{-1}$.
A possible physical explanation for the disappearance of the
soft excess may be a cooling of the disc temperature profile, which
shifts ${\rm T_{max}}$ out of the BeppoSAX sensitive bandpass.
Alternatively,
the lack of soft excess detection could be simply due to the improved
continuum determination, made possible by the broadband BeppoSAX
spectral coverage.
The 2--6~keV spectrum, against which Santos Lle\'o et al. (1994)
measured the soft excess in EXOSAT data, was flatter than in our
BeppoSAX baseline model (${\rm < \Gamma_{EXOSAT} \simeq 1.73 >}$).
If one fits the LECS/MECS data between
2 and 6~keV, one gets indeed ${\rm \Gamma \simeq 1.76}$. Extrapolating
this power-law in the 0.1--2~keV BeppoSAX/LECS leaves a complex residual
spectrum, with positive and negative wiggles, dominated by the
O{\sc vii} absorption feature. The interpretation of this residual
spectrum in the broad EXOSAT/LE filters is not totally
unambiguous.
We conclude that
the claimed soft
excess might be simply an artifact, due to extrapolating the
2--6~keV spectrum into the low energy filter EXOSAT
energy bandpass and/or to fitting a complex ionized absorber with the
Galactic cold photoabsorption only. On the other hand,
the lack of soft X-ray
excess is in line with the lack of blue bump
in this source (Santos-Lle\'o et al. 1994).

Another new result emerging from the BeppoSAX observation
of NGC~4593 is the
lack of any significant cutoff on the primary power-law, a
lower limit on ${\rm E_{cutoff}}$ being $\simeq 150$~keV.
This outcome is consistent with the measures of cutoff energies on
individual sources available so far: NGC~4151 (70--270~keV,
Zdziarski et al. 1995; Piro et al. 1998), IC4329A
(240--900~keV Madejski et al. 1995), MCG-6-30-15
(100-390~keV; Guainazzi et al. 1999), and NGC~5548 (110-200~keV; Nicastro
et al. 1999).

\begin{acknowledgements}
  
The BeppoSAX satellite is a joint Italian--Dutch program.
The authors wish to thank T.Mineo and A.Orr for useful discussions and
M.Salvati and G.Stirpe for comments on the manuscript. The anonymous
referee's comments helped us to better focus several issues.
MG acknowledges the receipt of an ESA Research Fellowship.
Financial support from ASI and CNR is acknowledged.
This research has made use of data obtained through
the High Energy Astrophysics Science Archive
Research Center Online Service, provided by the
NASA/Goddard Space Flight Center and of the NASA/IPAC Extragalactic
Database, which is operated by the Jet Propulsion Laboratory Caltech,
under contract with NASA.

\end{acknowledgements}

\end{document}